\documentclass[conference, a4paper]{IEEEtran}

\IEEEoverridecommandlockouts
\usepackage{graphicx,color}
\usepackage{amssymb,amsmath,times,wasysym,comment}
\usepackage{cite}
\usepackage{epsfig}

\title{\LARGE Broadband Non-Geostationary Satellite Communication Systems:\\ Research Challenges and Key Opportunities}

 \author{\normalsize Hayder Al-Hraishawi, Symeon Chatzinotas, and Björn Ottersten\\
	Interdisciplinary Centre for Security, Reliability and Trust (SnT), University of Luxembourg\\ 
} 

\begin{document}
	\maketitle
		
\begin{abstract}
	Besides conventional geostationary (GSO) satellite broadband communication services, non-geostationary (NGSO) satellites are envisioned to support various new communication use cases from countless industries. These new scenarios bring many unprecedented challenges that will be discussed in this paper alongside with several potential future research opportunities. 
	NGSO systems are known for various advantages, including their important features of low cost, lower propagation delay, smaller size, and lower losses in comparison to GSO satellites. However, there are still many deployment challenges to be tackled to ensure seamless integration not only with GSO systems but also with terrestrial networks.
	In this paper, we discuss several key challenges including satellite constellation and architecture designs, coexistence with GSO systems in terms of spectrum access and regulatory issues, resource management algorithms, and NGSO networking requirements. Additionally, the latest progress in provisioning secure communication via NGSO systems is discussed. Finally, this paper identifies multiple important open issues and research directions to inspire further studies towards the next generation of satellite networks.
\end{abstract}

%\begin{IEEEkeywords}
%	NGSO satellites, satellite communications, wireless security.
%\end{IEEEkeywords}

\section{Introduction}\label{sec:intro}
Satellites have a distinctive ability of covering wide geographical areas through a minimum amount of infrastructure on the ground, which qualifies them to be an appealing  solution to fulfill the growing diversified demand for broadband services \cite{Perez2019}. Currently, an increasing attention has been paid to the field of satellite communications from the global telecommunications market as several network operators start using satellites in the backhaul infrastructures for broadband connectivity and for the integration with 5G and beyond systems \cite{Giambene2018}. Recently, due to the swift rise of  “NewSpace” industries that are developing small satellites with new low-cost launchers, a large number of satellite operators are planning to launch thousands of non-geostationary (NGSO) satellites to satisfy the burgeoning demand for global high-speed and low-latency Internet connections \cite{Kodheli2020}. 
For instance, the emerging NGSO mega constellations such as OneWeb, Telesat, and Starlink have a system capacity reaching the
terabits-per-second level \cite{Su2019}.

NGSO satellites on a geocentric orbit include the low earth orbit (LEO), medium earth orbit (MEO) and highly elliptical orbit (HEO) satellites, which are orbiting constantly at a lower altitude than that of geostationary (GSO) satellites, and thus, their link losses are less and the latency due to signal propagation is lower \cite{ITU2003}. 
These intrinsic features of NGSO systems besides the high capacities, large footprints, and fast deployment, offer an interesting set of advantages for the  high-speed interactive broadband services that is even
competitive to terrestrial networks \cite{Guidotti2019}. Furthermore, the most newly developments in NGSO systems empower satellites to manage narrow steerable beams covering a relatively broad area, which facilities the use of smaller and lower cost equipment at the user terminals \cite{Guan2019,Hayder2020}. 
Specifically, the offered capacities by NGSO satellites can be further increased through utilizing high frequencies along with employing throughput enhancement techniques such as spatial diversity, smart gateway, multiple antenna at both user terminals and satellites, and  multiple-input multiple-output (MIMO) communications \cite{You2020}. 

These key benefits of NGSO satellites will increase the densification of their deployments, which might deteriorate the inter-satellite coexistence due to the  increased level of interference in the shared bands \cite{Hassan2020}. Therefore, understanding the interactions between the heterogeneous NGSO satellite systems is crucial to ensure seamless coexistence.
Additionally, the emerging NGSO services are expected to operate in the same frequency bands of the existing GSO systems, which will threaten their coexistence in these bands and it will be one of the challenging issues that will require developing novel satellite interference mitigation techniques or probably some regulatory interventions \cite{Braun2019}. 
Moreover, due to the highly dynamic feature of NGSO constellations and spatially heterogeneous traffic demands, the scenario of having heavily loaded satellite links while others are underutilized will inevitably occur frequently, and eventually leads to buffer overflows, higher queuing delays, and significant packet drops at the congested satellites \cite{te_hayder2020}. Hence, developing load-balancing algorithms to guarantee a better traffic distribution among satellites is indispensable.

Notwithstanding the growing interest in NGSO satellites due to their essential features  that can be envisaged for high-speed interactive broadband services, there are still many brand new challenges in the NGSO system evolution to be addressed  to achieve high quality communications. 
The objective of this paper is to investigate the related forward-looking challenges of NGSO systems development and integration with highlighting  the applications targeted by NGSO communication systems. By this we aim to discuss the following key implementation aspects:
\begin{itemize}
	\item	It is of paramount importance to study communication system design and rearchitecturing directions to integrate NGSO with terrestrial networks to ensure seamless global coverage consistent with satellite constellation design.
	
	\item	NGSO systems have also to confront the interference issue due to the coexistence with other satellite systems, and thus, developing novel interference coordination/mitigation techniques is crucial.  
	\item The heterogeneity of NGSO systems alongside with the relative movement between satellites in the lower orbits may affect the system performance. Thereby, resources management mechanisms are indispensable in such dynamic propagation environments 
	
	\item The growing number of NGSO satellites increases system complexity and leads to establish space-based networks to improve coordination and resource utilization. Thus, introducing inter-satellite communication links is inevitable in such setups.
	
	\item The integration of NGSO satellite systems into Internet infrastructures comes with serious security threats due to the large constellations that will include hundreds or even thousands of satellites providing direct connectivity. 
\end{itemize} 
Beyond this, several prospective future research directions and opportunities for NGSO systems are also provided.

\section{Open Challenges}
Despite the potential advantages offered by NGSO satellite communication systems, open essential challenges still need to be addressed.

\subsection{NGSO Satellite Constellation Design}
Generally, satellite orbit constellation design  is a key factor that directly affects the performance of the entire satellite systems. The fundamental constellation parameters include the type of orbit, altitude of the orbit, number of orbits, number of satellites in each orbit, and satellite phase factor between different orbit planes \cite{Qu2017}. Several earlier studies have considered systematic constellation patterns of satellites such as polar constellations and Walker-Delta patterns \cite{Walker1984}, which are formulated based on the relative positions of the satellites in the earth-centered inertial frame (ECI). Additionally, in \cite{Mortari2004}, the concept of flower constellations has been proposed to put all satellites in the same 3D trajectory in the earth-centered earth-fixed frame (ECEF). However, these design approaches do not take into consideration the demand characteristics on earth, which makes them inefficient  strategies when bearing in mind the non-uniform and uncertain demand over the globe. Thus, a more competent strategy would be a staged flexible deployment that adapts the system to the demand evolution and begins covering the regions that have high-anticipated demands.

Another relevant constellation concept that can be applied to NGSO systems was proposed in \cite{Paek2012} to constitute reconfigurable satellite constellations where satellites can change their orbital characteristics to adjust global and regional observation performance. This concept allows establishing flexible constellation for different areas of interest. However, introducing reconfigurability feature to the constellation requires a higher maneuvering capability of the satellites and more energy consumption and that can be a deterrent factor when multiple successive reconfigurations are needed over the life cycle. On the other hand, a hybrid constellation design is proposed in \cite{Chan2004} to utilize multiple layers and mixed circular-elliptical orbits, and thus, accommodating the asymmetry and heterogeneity of the traffic demand. Nonetheless, the optimization of adapting the constellation to growing demand areas is a challenging issue to be addressed in the context of integration an entire hybrid model. 
Moreover, an integrated framework accounts for the spatial-temporal traffic distributions and optimizes the expected life cycle cost over multiple potential scenarios can be an initial plan to circumvent the NGSO constellation design challenges.

Furthermore, traditional global-based constellation systems are no longer valid solutions for NGSO systems due to high cost and inflexibility to react to uncertainties resulting from market demands and administrative issues. Therefore, regional coverage constellations are promising solutions for satellite operators as they will be able to tackle the economic and technical issues in a flexible manner \cite{Lee2020}. Regional constellations focus on the coverage over a certain geographical region by using a small number of satellites in the system and they can achieve the same or better performance compared to global-coverage constellations.
Regional coverage constellations can also provide sufficient redundancy with deploying  multiple NGSO satellites in lieu of a single GSO satellite, and thus, operators can hand off traffic to satellites that avoid beam overlapping,  and therefore avoid interference \cite{LEE2018213}.
However, designing an optimal regional constellation is a complicated process, which requires optimizing the orbital characteristics (e.g., altitude, inclination) while considering asymmetric constellation patterns, particularly for complex time-varying and spatially-varying coverage requirements. This topic has not been deeply investigated in the literature, and thus, new sophisticated approaches to design optimal constellation patterns are needed to be developed and  tailored to different orbital characteristics and NGSO environments.

\subsection{Coexistence of NGSO and GSO satellites}
According to the ITU regulations, the interference inflicted at GSO satellites from NGSO satellite systems shall not degrade GSO satellites performance and shall not claim protection from GSO systems in the fixed-satellite and broadcasting-satellite services \cite{ITU2003}. Specifically, the effective power flux density (EPFD) within the frequency bands that are allocated to GSO systems and at any point on the earth’s surface visible from the GSO satellite orbit shall not exceed the given predefined limits in the ITU regulations. 
Although NGSO systems have potentials of global coverage and high performance, many of their regulatory rules were coined nearly two decades ago based on the proposed technical characteristics of NGSO satellites at the time. This is very challenging from a spectral coexistence viewpoint, and it will require much more agile systems.
Moreover, the deployment of NGSO satellites is undergoing a significant densification comparing to existing GSO systems, which is leading to unprecedented inter-satellite coexistence challenges. The high interference levels will not only resulting from the enormous number of operating satellites but also due to the expected high heterogeneity of the NGSO systems. Therefore, it is imperative to scrutinize the interference interactions between different GSO and NGSO systems to ensure consistent hybrid deployment landscape.

Despite the several prior works on developing interference mitigation techniques for satellite systems, the high heterogeneity and ambiguity about the parameters of the emerging deployments make the effectiveness of these traditional mitigation techniques questionable. Moreover, most of the prior works focus mainly on the inter-system interference between GSO and NGSO, while the serious issue of NGSO-NGSO interference was recently  addressed only in \cite{Braun2019} and \cite{Tonkin2018}.
Specifically, the inter-satellite coexistence in the Ku-band is investigated in \cite{Braun2019} by studying the impact of both NGSO-NGSO and NGSO-GSO co-channel interference on throughput.
The impact of NGSO-NGSO co-channel interference on the achievable throughput for NGSO constellations is studied in \cite{Tonkin2018}. Band splitting interference mitigation techniques are also investigated in \cite{Tonkin2018} with considering the Ka and V bands. Accordingly, the highly heterogeneous NGSO constellation properties and GSO-NGSO interference interactions need to be thoroughly analyzed for satellite deployments over different bands and constellations.

The concept of mega-constellation brings about spectrum sharing challenges between NGSO and GSO systems. These mega-constellation satellites will operate at the same frequencies that are currently used by GSO satellites including the Ka and Ku bands, which has raised some serious concerns among GSO satellite operators \cite{Giambene2018}. Therefore, coordination and awareness of the operational characteristics about each counterpart system is essential in order to achieve a successful spectrum sharing between different satellites. A database-based operation is foreseen a possible approach can achieve sort of coordination between mixed satellite systems \cite{Hoyhtya2017}. Additionally, cognitive satellites with spectrum sensing and awareness can be used in this context along with resource allocation techniques, i.e., carrier power and bandwidth \cite{Lagunas2015}. For instance, designing a cognitive spectrum utilization scenario could circumvent the coexistence challenges, where NGSO satellites can exploit the spectrum allocated to GSO satellites or terrestrial networks without imposing detrimental interference to their concurrent transmissions.

 \begin{figure}[!t]
	\centering
	\def\svgwidth{210pt}
	\fontsize{8}{4}\selectfont
	\scalebox{1}{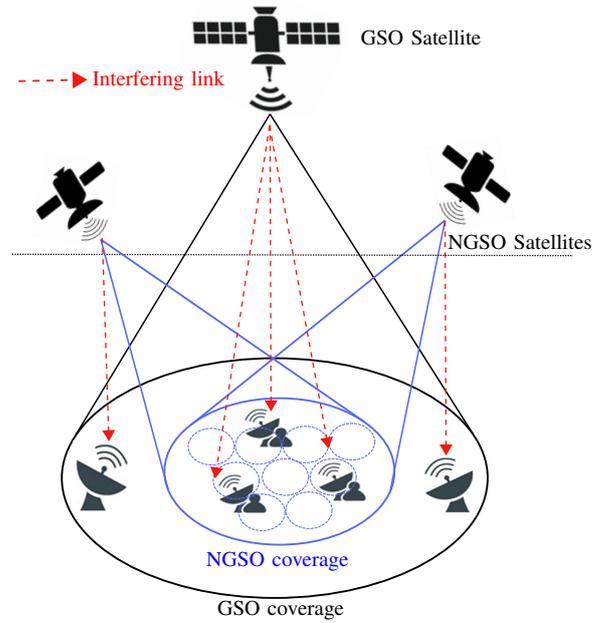}
	\caption{Interference scenario involving a GSO and multiple NGSO satellites.} \label{fig:system_model}
\end{figure}

Interference analysis of the emerging NGSO constellations should take into consideration the effect of the aggregated interference due to utilizing a large number of multi-beam satellites and applying frequency reuse techniques.
Fig. \ref{fig:system_model} shows an interference scenario where multiple satellites having multi-beam and multi-carrier per beam. Interference to noise ratio (I/N) is an important metric in interference evaluation, and it can be calculated for this scenario as follows
\begin{eqnarray}
	I/N=\frac{\sum_{i=1}^{S}\sum_{j=1}^{L} I_{i,j}	}	{K B_d T_d}
\end{eqnarray}
where $I_{i,j}$ represents the downlink single entry interference level at the users on ground from the $j$-th link of the $i$-th satellite, $I_{i,j}$ can be computed as given in \cite{Wang2020}. $S$ denotes the number of interfering satellites and $L$ denotes the number of interfering links. $K$ accounts for Boltzmann constant, $B_d$ represent the downlink bandwidth, and $T_d$ is is the receive noise temperature of the receive antenna.  
In this context, power control techniques for both the uplink and the downlink can improve the coexistence performance by mitigating the inter-satellite interference. For example, minimizing the downlink transmit power $P_{t_i}$ of the NGSO satellites can be formulated as an optimization problem as follows:
\begin{subequations}
	\begin{eqnarray}
		&& \mathop{\text{minimize}}\limits_{d_{ss} \; \forall S}  \hspace{4mm}  P_{t_{i}} \\
		&& \text{subject to} \hspace{4mm} C_i/N_i \geq C_0/N_0 , \;\; \forall S, \\
		&& \hspace{18mm} I_{i,NGSO} \leq I_{th}, \;\; \forall S,
	\end{eqnarray}
\end{subequations}

\noindent
where $d_{ss}$  is the distance corresponding to downlink interference path with other satellites. $C_i/N_i$ represents the carrier to noise ratio of the $i$-th satellite at the receiver and it should be greater than $C_0/N_0$ to keep the link transmitting. Moreover, GSO links are protected against the concurrent NGSO  transmissions by maintaining the detrimental co-channel interference  to an acceptable level that is less than a predefined interference threshold ($I_{th}$).

\subsection{Resource Management and Orchestration}
The high complexity and variety of NGSO satellite architectures and the high-speed mobility with respect to the earth’s surface inflict multiple resource management challenges that need to be carefully addressed. 
The NGSO satellites are moving in a higher speed than GSO systems causing more frequent handovers between satellites during the service \cite{Li2019}. Further, the spectrum allocated for the applications that served by NGSO systems is neither constant nor fully dedicated during the service interval. Specifically, the spectrum resource blocks are allocated based on the available spectrum resources, the speed requirement, and the priority of the service and user. 
These enhanced and extended features of NGSO satellites with their heterogeneous resources are exacerbating the resource management challenges. Thus, integrated approaches for spectrum access strategies that are cognizant of the advantages of different satellite systems are needed to be  flexible and highly adaptable.

Resource management is significantly affected by the employed satellite coverage scheme, where in this regards there are  two popular coverage schemes can be adopted by NGSO systems: (i) spot beam coverage and (ii) hybrid wide-spot beam coverage \cite{Lutz2012}. In a spot beam coverage scheme, each satellite provides multiple spot beams to offer coverage over its service area, where their footprint on earth's surface moves along with the satellite trajectory. This scheme is simple but the handovers between beams are more frequent because the coverage area of a single spot beam is comparatively small.
On the other hand, in hybrid wide-spot beam scheme each satellite provides a wide beam for the whole service area and several steering beams for users employing digital beamforming techniques. The spot beams are always steered to the users, and thus, the provided footprint is nearly fixed during the movement of satellite. In this scheme, handovers occur between the wide beams only of any two satellites, and then, the number of handovers substantially decreases due to the vast coverage of the wide beam.

Thereby, the performance of NGSO systems should be optimized based on its resource utilization, namely, time, power, spectrum, antenna, satellite beams, and orbital planes. When the available NGSO resources are combined with users and services management over different requirements, several interesting resource optimization opportunities arise. Nevertheless, formulating and solving such problems become more complicated because of the broad resource pool, the interference issues, and channel state information (CSI) availability challenges. Additionally, resource management schemes to support different satellite systems can be designed and tailored to various objectives, e.g., maximizing the achievable throughput, minimizing the consumed power,  reducing latency, and better quality of service. In general, development of advanced resource allocation algorithms under different resource availability in the coexistence of NGSO and GSO systems persists to be an intriguing research direction.

\subsection{NGSO Networking}
Currently satellite systems are witnessing a paradigm shift due to the rise of NGSO satellites compared to the existed GSO satellites. Specifically, GSO systems are in constant contact with ground stations as they control the GSO operations, whereas NGSO systems will need to be built on more autonomous and reconfigurable architectures, and the assumption of persistent contact with ground stations is infeasible in the NGSO case. Thus, NGSO satellites can be deployed as a space-based network using inter-satellite links, which tends to be the mainstream of the ongoing NGSO developments. 
Space-based networks can fulfill the increasing complexity of application requirements. Establishing space-based network architectures is more economically efficient and more suitable for the heterogeneous integrated satellite communications. 
The inter-satellite links in space-based networks eliminate the use of the excessive number of gateways. This architecture is particularly favorable for the areas where acquiring gateway sites is difficult \cite{Hassan2020}.
Additionally, space-based networks enable communication and cooperation between satellites for traffic routing, throughput maximization, latency minimization, and seamless coverage. These networks have been already used in some applications, such as navigation, positioning, and weather forecasting satellites, and then, they are anticipated to play a profound role in NGSO satellite networks and the global communication infrastructures in general.

The expected better performance of space-based networks will be achieved at the cost of higher complexity that is necessary for load balancing between satellite links and for finding paths with the shortest end-to-end propagation delay, which imposes some restrictions on deploying NGSO networks for delay-sensitive applications. 
Given the high-mobility of NGSO satellites (especially LEO systems) with respect to the slow moving or fixed user terminals as well as the arbitrary traffic loads, many concerns are raised about current mobility management mechanisms and routing algorithms and their efficiency to tackle these limitations. 
Specifically, disregarding the optimization aspects between traffic distribution and routing strategies may lead to considerable queuing delays and increased number of packet drops at the heavily-loaded satellite links \cite{Radhakrishnan2016}.
Inter-satellite route planning to satisfy the required quality of service levels among users is quite challenging for the development of NGSO satellite networking, especially when employing a large number of satellites and network topology is constantly changing. Besides, many parameters are involved for determining the optimal path such as delay, bandwidth, path reliability, link status, traffic load, hop count, etc. Thus, it is essential to design novel routing protocols  that are able to find the optimal route for a data packet to be transmitted between a sender and a receiver while taking into consideration the mobility of satellite nodes.

\subsection{Security Challenges}
In this subsection, we discuss the security challenges that need to be addressed in order to make NGSO satellite systems more secure while maintaining seamless interaction with GSO systems and terrestrial networks. Generally, satellite communications have been relied on terrestrial base stations for provisioning secured transmissions, which pushed the majority of security research efforts to focus mainly on the data links between satellites and the terrestrial base stations, i.e., uplink and downlink \cite{He2019}. However, the steadily growing deployment of the space-based wireless network shows that there will be also a big security risk in the data communication between satellites and even the internal structure of satellites. These security issues cannot be ignored and they deserve more attention.  Additionally, the complex structure of the space-based wireless network requires various security modeling and analysis for the space-based NGSO networks in combination with certain application scenarios.

Proper security mechanisms are essential for NGSO communication systems because they are susceptible to security threats such as eavesdropping, jamming, and spoofing. For instance, any sufficiently well-equipped adversary can send spurious commands to the satellite and gain full access to satellites as well as data, enabling them to cause serious damage. 

Another example when satellite systems use strong security mechanisms for performing message-integrity checks or authenticating users, denial-of-service attacks can be conducted by adversaries via sending a large number of spurious messages to the satellite  \cite{Chowdhury2005}. Thus, satellites under this attack will spend significant computational  processing power and time to the spurious messages, which degrades the quality of service for the legitimate users. NGSO satellites can be particularly susceptible to this kind of attacks because it is a single point of failure and can be easily overwhelmed if compel to execute excessive computations.

Security of satellite communication is traditionally provisioned through cryptographic-based techniques on the upper layers, which requires high computational complexity \cite{Xiao2019}.
Towards this end, on one hand, free space optical (FSO) communication technology is an interesting alternative to RF inter-satellite-links owing to the wide bandwidth and high data rate that an FSO system can offer, where the optical technologies are foreseen as a key enabler for ultra-secure communications with the use of, e.g., quantum key distribution \cite{Podmore2019}.
 
On the other hand, as a complementary technique to the traditional cryptographic-based methods, physical layer security has been proven to be an effective approach to achieve everlasting security without the heavily computational processes of encryption/decryption. Physical-layer security technique can be introduced as an added layer of defense into NGSO satellites, but the studies in this area are still nascent.

\section{Potential Opportunities}
In addition to the key challenges  that have been raised in the previous section about NGSO network architectures and designs, this section briefly discusses some innovative research directions and new opportunities for utilizing NGSO systems to realize advanced satellite communications for versatile applications.

\subsection{Space-based Cloud}
Far from the common use of satellites as relay devices, the space-based cloud concept has emerged as a promising and secured paradigm for data storage over NGSO satellites, particularly in the context of big data technologies and applications \cite{Jia2017}. The key advantage of space-based data storage is providing complete immunity from natural disasters occurring on Earth. Furthermore, utilizing NGSO satellites for data storage can offer more flexibility to some cloud networks that designed to transfer data globally regardless the geographical boundaries and terrestrial obstacles \cite{Huang2018}. For instance, mega-corporations and large organizations that are located at different global sites can share big data through a space-based cloud and benefit from the faster transfer rate comparing to the traditional terrestrial cloud networks, especially for delay-sensitive services. Thereby, NGSO satellites could expand their scope of missions for more than only operating as relay devices for communication networks.

\subsection{IoT via NGSO Satellites}
The flexibility and scalability properties of NGSO satellites make their employment within the Internet of Things (IoT) ecosystem more appealing to shape novel architectures that uplift the interoperability among a plethora of applications and services \cite{Qu2017}. Thus, by exploiting the relatively shorter propagation distances of NGSO satellite constellations, IoT terminals can be designed to be small-sized, long-life, and low power consumption, which achieves IoT ideal operation schemes. Moreover, the reduced operating expenditures (OPEX) and capital expenditures (CAPEX) of NGSO satellites comparing to GSO ones render them into a more feasible way to deploy efficient IoT services over wide geographical areas \cite{Bacco2018}. Hence, these exceptional features of NGSO satellites can unleash the full potentials of IoT, and that will establish a universal network of billions of interconnected devices.
However, there are many technical challenges in connecting NGSO satellites to mobile or stationary devices, and this task particularly requires  a unified network vision on providing hybrid connectivity and prototyping satellite technology to support the advancement in machine-to-machine communications and IoT technologies.

\subsection{Caching Over NGSO Satellites}
Benefiting from the high-capacity backhaul links and ubiquitous coverage, NGSO satellites can help bring content closer to the end users, and thus, these satellite can be considered as an option for data caching. NGSO satellites also have the ability to multi-cast data and quickly update the cached content over different locations \cite{Liu2018}. Additionally, the symbiotic relationship between satellite and terrestrial telecommunication systems can be exploited to create a hybrid federated content delivery network, which will substantially ameliorate user experience \cite{Vu2018}.
Therefore, integration of NGSO satellites into future Internet with enabling in-network caching makes traffic demands from users for the same content to be easily accommodated without multiple transmissions, and thereby, more spectral resources can be saved along with reducing transmission delay.
However, the time-varying network topology and limited on-board resources in NGSO satellites have to be taken into account when designing caching placement algorithms alongside with their fast convergence and low complexity.

\subsection{Terahertz Communications}
Terahertz (THz) band communications are anticipated to support a wide variety of applications in the upcoming 6G wireless networks \cite{Akyildiz2020}. Besides the identified potential indoor THz use cases and scenarios, there are some foreseen THz communication applications within the context of the integrated space information networks, e.g., the satellite cluster networks and inter-satellite backbone networks \cite{Suen2015}. Unlike ground THz communications that suffer from short distance transmission limitations due to the atmosphere attenuation, deploying THz communications in space applications in the atmosphere-free environment
circumvents this constraint and achieves high-speed long-distance links between satellites. However, there are still a number of open challenges for THz satellite communications particularity in terms of semiconductor technologies. For example, it is prohibitively difficult to produce high power THz transmitters and current THz receivers prone to higher noise figures. 
Thereby, with more research efforts dedicated for developments of high power THz transmitters, highly sensitive receivers, and adaptive antenna arrays, many THz communication opportunities can be explored within the NGSO satellite deployments.

\section{Conclusions}\label{sec:conclusions}
This paper brings up several technical challenges and thought-provoking research directions to integrate NGSO satellites into the global wireless communication platforms. First, the most foreseen satellite constellation design challenges and rearchitecturing directions are explored to ensure seamless coverage with considering  GSO satellites and terrestrial networks. Then, the coexistence challenges in terms of interference and spectrum access of NGSO integration within the ecosystems of GSO satellites and terrestrial networks are discussed. The diverse NGSO resources and their management algorithms are highlighted as well. Afterwards, the new requirements for NGSO networking and the associated security threats are deliberated. 
Furthermore, new application scenarios of the NGSO satellites with novel features and properties are presented. In short, this paper does not attempt to address every aspect of NGSO satellites but we are hopeful that it would trigger some more in-depth thinking and enrich the state-of-the-art of  NGSO communication systems.

\section*{Acknowledgement}
This work is financially supported by the Luxembourg National Research Fund (FNR) under the project MegaLEO (C20/IS/14767486).

\linespread{1.2}
% ===========================================================================
% bibliography
% ===========================================================================

\bibliographystyle{IEEEtran}
\bibliography{IEEEabrv,References}

\end{document}